\def\mK{{\rm \mu K}}

\def\expec#1{\langle#1\rangle}

\def\etal{{\frenchspacing\it et al.}}
\def\ie{{\frenchspacing\it i.e.}}
\def\eg{{\frenchspacing\it e.g.}}
\def\etc{{\frenchspacing\it etc.}}
\def\rms{{\frenchspacing r.m.s.}}

\def\pp{\noindent\parshape 2 0truecm 13.6truecm 1truecm 12.6truecm}
\def\rf#1;#2;#3;#4 {\par\pp#1, {\it #2}, {\bf #3}, #4. \par}
\def\rg#1;#2;#3;#4;#5 {\par\pp#1, {\it #2}, {\bf #3}, #4 (``#5"). \par}
\def\rn{\pp}

\def\beq#1{\begin{equation}\label{#1}}
\def\eeq{\end{equation}}
\def\beqa#1{\begin{eqnarray}\label{#1}}
\def\eeqa{\end{eqnarray}}
\def\eq#1{equation~(\ref{#1})}
\def\eqnum#1{~(\ref{#1})}

\def\bfig{\begin{figure}[h] \centerline{\hbox{}}\vfill}
\def\efig{\end{figure}\vfill\newpage}

\def\fheight{12cm}
\def\fwidth{17cm}
\def\fig#1{Figure~\ref{#1}}

\def\spose#1{\hbox to 0pt{#1\hss}}
\def\simlt{\mathrel{\spose{\lower 3pt\hbox{$\mathchar"218$}}
     \raise 2.0pt\hbox{$\mathchar"13C$}}}
\def\simgt{\mathrel{\spose{\lower 3pt\hbox{$\mathchar"218$}}
     \raise 2.0pt\hbox{$\mathchar"13E$}}}
\def\simpropto{\mathrel{\spose{\lower 3pt\hbox{$\mathchar"218$}}
     \raise 2.0pt\hbox{$\propto$}}}

\def\addr#1{{\small\it #1}}

\def\n{\varepsilon}
\def\vr{{\bf r}}
\def\vn{{\bf n}}
\def\tr{\hbox{tr}\>}
\def\P{{\bf P}}

\def\ed{\end{document}}

\documentstyle[11pt,epsf,rotate]{article}
\begin{document}

%%%%%%%%%%%%%%%%%%%%%%%%%%%%%

\begin{titlepage}   % Not numbered.

\noindent
%\today
%\hfill MPI-PhT/96-67
\begin{center}

\vskip0.9truecm
{\bf

A HIGH-RESOLUTION MAP OF THE COSMIC MICROWAVE BACKGROUND AROUND
THE NORTH CELESTIAL POLE\footnote{
Published in {\it ApJ Letters}, {\bf 474}, L77. Available from \\
{\it h t t p://www.mpa-garching.mpg.de/$\tilde{~}$max/saskmap.html} (faster from Europe)\\
and from {\it h t t p://www.sns.ias.edu/$\tilde{~}$max/saskmap.html} (faster from the US).\\
Note that the figures will print in color if your printer supports it.}
}

\vskip 0.5truecm

Max Tegmark$^{1,2,3}$, 
Ang\'elica de Oliveira-Costa$^{2,3,4}$, 
Marc. J. Devlin$^4$, 
C. Barth Netterfield$^4$, 
Lyman Page$^4$ \& 
E. J. Wollack$^4$

\smallskip
\addr{$^1$Max-Planck-Institut f\"ur Physik, 
F\"ohringer Ring 6,D-80805 M\"unchen}\\
\addr{$^2$Max-Planck-Institut f\"ur Astrophysik, 
%: Karl-Schwarzschild-Str. 1, D-85740 Garching}\\
Karl-Schwarzschild-Str. 1, D-85740 Garching; angelica@mpa-garching.mpg.de}\\
\addr{$^3$Institute for Advanced Study, 
Olden Lane, Princeton, NJ 08540; max@ias.edu}\\
\addr{$^4$Princeton University, Department of Physics, Jadwin Hall, 
Princeton, NJ 08544; page@pupgg.princeton.edu}\\

\vskip 0.7truecm

\end{center}

\abstract{
We present a Wiener filtered map of the Cosmic Microwave Background (CMB)
fluctuations in a cap with $15^\circ$ diameter,
centered at the North Celestial Pole. 
The map is based on the 1993-1995 data from the Saskatoon (SK) experiment,
with an angular resolution around $1^\circ$ in the frequency 
range 27.6--44.1 GHz. The signal-to-noise ratio in the map 
is of order two, and some individual hot and cold spots are
significant at the $5\sigma$-level.
The spatial features are found to be consistent from year to year, 
reenforcing
the conclusion that the SK results are not dominated by
residual atmospheric contamination or other non-celestial signals.
}

\end{titlepage}

%%%%%%%%%%%%%%%%%%%%%%%%%%%%%%%%%%%%%%%%%%%%%%
\def\m{m}	% Number of pixels
\def\n{n}	% Number of data points
\def\pix{\phi}	% Pixel profile (typically square or Gaussian)
\def\sumi{\sum_{i=1}^{\m}}
\def\ith{i^{th}}
\def\f{f}
\def\x{x}
\def\y{y}
\def\n{n}
\def\vx{{\bf x}}
\def\vxt{\tilde{\bf x}}
\def\vy{{\bf y}}
\def\rh{\widehat{\bf r}}
\def\F{{\bf F}}
\def\I{{\bf I}}
\def\N{{\bf N}}
\def\S{{\bf S}}
\def\W{{\bf W}}

% Boldface Greek letters:
\font\bfmath=cmmib10
\def\bfvarepsilon{\hbox{\bfmath\char'042}}
% See p. 430 of the TeXBook for the other character numbers.
\def\ve{\bfvarepsilon}

\section{INTRODUCTION}

Since the fluctuations in the Cosmic Microwave Background (CMB)
depend on a large number of cosmological parameters (see 
Hu, Sugiyama \& Silk 1996 for a recent review), 
accurate CMB measurements 
could enable us to measure parameters such as the Hubble constant, the
density parameter $\Omega$, {\etc} to hitherto unprecedented accuracy
(Jungman {\etal} 1996). 
After the successful measurements of large-scale fluctuations
by COBE team (Smoot {\etal} 1992; Bennett {\etal} 1996), attention is now shifting towards
measurements at higher angular resolution. 

When reducing a CMB data set, one usually wants to produce either 
an estimate of the angular power spectrum $C_\l$ or a map.
Although it is the former that is ultimately used to constrain
cosmological parameters, there are a number of reasons for why map making 
is useful as well (apart from a general desire to map the
sky in as many frequency bands as possible):
\begin{itemize}
\item
It facilitates comparison with other experiments.

\item
It facilitates comparison with foreground templates such as the DIRBE maps.

\item
It may reveal flaws in the model that are not visible in the power spectrum,
such as non-Gaussian CMB features, point sources and spatially localized 
systematic problems.
\end{itemize}
%As high-resolution experiments increase in number, 
%it is therefore desirable to produce high-resolution maps
%which can help increase our understanding of foregrounds, 
%systematics {\etc} on these scales. 
The first degree-scale map reconstructed from difference measurements
was produced by White \& Bunn (1995) using data
from the MAX experiment, which probed a strip of sky a few degrees wide 
at a resolution of half a degree.
The purpose of this {\it Letter} is to present a map based on data 
from the Saskatoon (SK) experiment
(Wollack {\etal} 1996; Netterfield {\etal} 1997, hereafter ``N97").
This map covers a larger patch of sky: a cap of $15^\circ$ diameter 
centered on the North Celestial Pole (NCP). 
The angular resolution is similar to that of MAX (about half a 
degree for the SK95 data), but the region in question is 
more evenly sampled than was the case with the MAX map, with no ``holes". 

The map-making method we employ 
is described in Section 2 and the results are presented and
discussed in Section 3.

\section{METHOD}

\begin{figure}[phbt]
\centerline{{\vbox{\epsfxsize=14cm\epsfysize=14cm\epsfbox{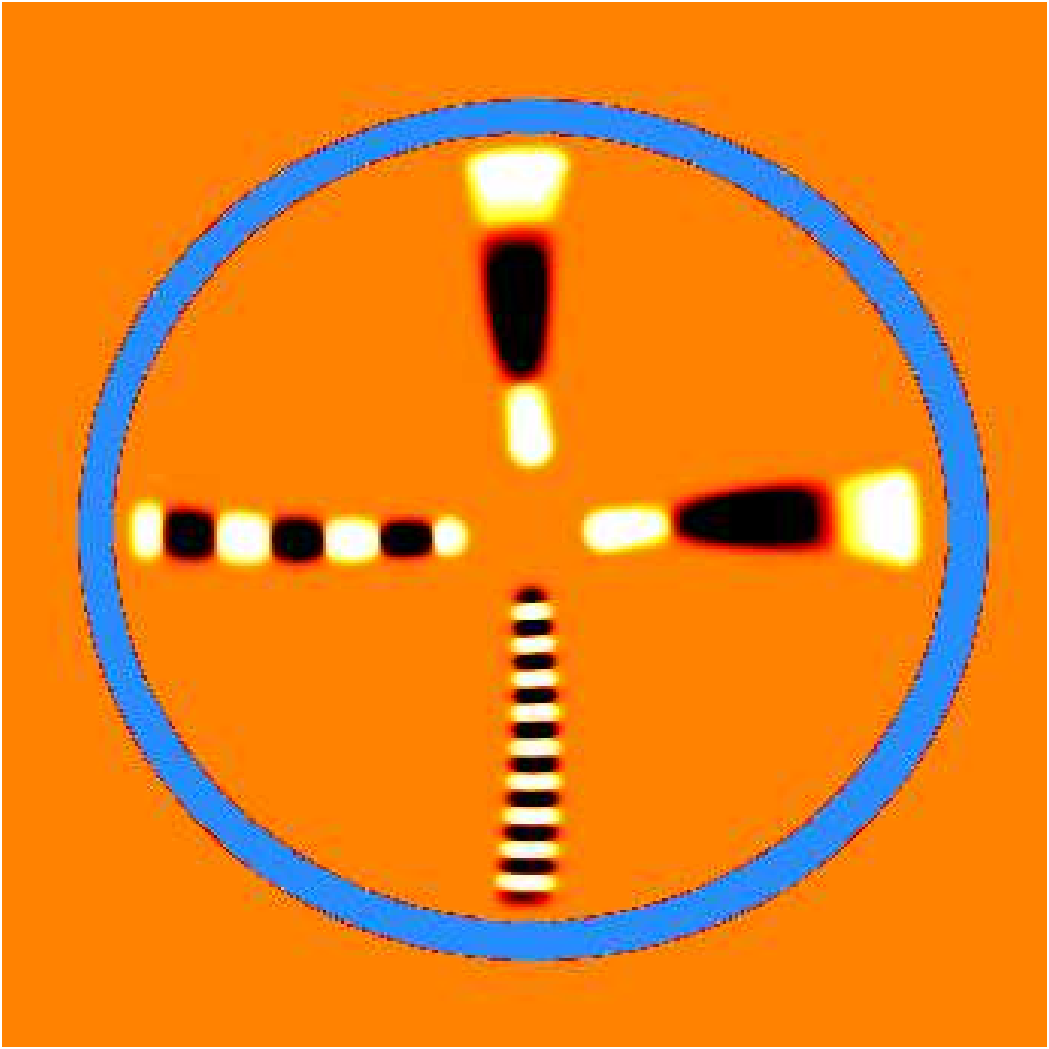}}}}
%\centerline{\rotate[r]{\vbox{\epsfysize=11cm\epsfysize=11cm\epsfbox{beamfig.ps}}}}
%\vskip-2cm
\caption{
Sample Saskatoon weight functions.
}
Four of the 2590 Saskatoon weight functions are shown in a circle 
of diameter $16^\circ$ with the North Celestial Pole at the center.
The weight functions are all for the 1995 East data, and correspond to
the 3-point beam in the $RA=$0h azimuthal bin (top),
the same beam in the $RA=$6h azimuthal bin (right),
the 19-point beam in the $RA=$12h azimuthal bin (bottom) and 
the 7-point beam in the $RA=$18h azimuthal bin (left). 
\label{WeightFig}
\end{figure}

The task of generating maps from the Saskatoon data is 
complicated by the fact that the data set does not contain 
simple sky temperatures, but rather $2590$ different 
linear combinations of sky temperatures\footnote{
The present analysis is based on the CAP 
data (as defined in N97) and does not include the RING data.
}. 
The $\ith$ data point, which we denote $y_i$, is a linear combination 
of the temperature across the sky, where the weights ascribed to each patch of 
sky are given by some known function $\f_i(\rh)$.
These weight functions are described in detail in N97. For illustration, 
four sample weight functions are shown in \fig{WeightFig}.
All weight functions emanate approximately radially from the NCP, 
oscillate in the radial direction, cover only a small 
RA-band, and extend to about $8^\circ$ from the pole.
The rest of this section describes the inversion process of reconstructing a map
from these linear combinations $\y_i$.

\subsection{Wiener filtering}

Wiener filtering is a general method for estimating a signal 
from noisy data (Wiener 1949),
and can be derived as follows.\footnote{
If the data set is Gaussian, an alternative
derivation the Wiener filter is to maximize 
the a posteriori probability distribution, {\ie}, to 
find the most likely map given the data 
(Zaroubi {\etal} 1995). 
}
Suppose that we have a vector of $\n$ data points $\vy$ and wish to estimate 
a vector of $\m$ numbers $\vx$ (for instance the pixel temperatures in a map) 
from it. 
Without loss of generality, we can assume that 
both vectors have zero mean, {\ie}, 
$\expec{\vx} = 0$ and $\expec{\vy} = 0$, since otherwise,
we could  redefine them so that they do.
Denoting the estimate of $\vx$ by $\vxt$, 
the most general linear estimate can 
clearly be written as 
\beq{ReconstructionEq}
\vxt\equiv\W\vy
\eeq
for some $\m\times\n$ matrix $\W$. 
Defining the error vector as 
$\ve\equiv\vxt-\vx$,
a natural measure of the errors is
the quantity $|\ve|$, which is just $\m$ times 
the {\rms} error per data point. 
The expectation value of $|\ve|^2$ is given by
\beq{ErrorEq1}
\expec{|\ve|^2} 
=\expec{(\W\vy-\vx)^t(\W\vy-\vx)}
= \tr\left[
\W\expec{\vy\vy^t}\W^t - 2\expec{\vx\vy^t}\W^t + \expec{\vx\vx^t}\right].
\eeq
The Wiener filter $\W$ is the matrix that minimizes this error.
\footnote{
Since the reconstruction of pixel $i$ depends only on the $i^{th}$ 
row of $\W$, this is equivalent to minimizing 
all the errors $\varepsilon_i^2$ separately.
}
By differentiating with respect to the components of $\W$, we obtain the 
simple result
\beq{WienerEq}
\W =  \expec{\vx\vy^t}\expec{\vy\vy^t}^{-1}.
\eeq
Direct substitution shows that the covariance matrix of the estimates is
\beq{EstCovEq}
\expec{\vxt\vxt^t} = \expec{\vx\vy^t}\expec{\vy\vy^t}^{-1}\expec{\vy\vx^t}
\eeq
and that the error covariance matrix is
\beq{ErrCovEq}
\expec{\ve\ve^t} = 
\expec{\vx\vx^t} - \expec{\vx\vy^t}\expec{\vy\vy^t}^{-1}\expec{\vy\vx^t}.
\eeq
Linear filtering techniques have recently been applied to a 
range of cosmological problems. Rybicki \& Press (1992) give a detailed 
discussion of the one-dimensional problem. 
Lahav {\etal} (1994), Fisher {\etal} (1995) and Zaroubi {\etal} (1995)
apply Wiener filtering to galaxy surveys. 
The COBE DMR maps have been processed both with Wiener filtering
(Bunn {\etal} 1994; Bunn {\etal} 1996) and with other linear filtering
techniques (Bond 1995).

\subsection{Application to the Saskatoon case}

In our case, the observed data point $\y_i$ is the true
sky temperature distribution
$\x(\rh)$ convolved with the $\ith$ beam function $\f_i(\rh)$, with noise 
$\n_i$ added
afterwards, so 
\beq{yDefEq}
y_i = \n_i + \int\f_i(\rh) x(\rh) d\Omega. 
\eeq
Our maps will cover a square region of $20^\circ\times 20^\circ$
centered on the NCP, pixelized into a $64\times 64$ 
square grid, so $\m=4096$.
To ensure that the maps are properly oversampled, we define 
the pixels to be the sky temperatures after Gaussian 
smoothing on a scale of $\sigma=1^\circ$:
\beq{xDefEq}
x_i = \int \pix(\rh_i\cdot\rh) x(\rh) d\Omega,
\eeq
where $\rh_i$ is a unit vector in the direction of the
$\ith$ pixel and 
\beq{GaussianEq}
\pix(\cos\theta) \equiv {1\over 2\pi\sigma^2}e^{-\theta^2/2\sigma^2}.
\eeq
This means that the map resolution $\sigma$ is 
$3.2$ times the pixel separation $20^\circ/64$, which
is safely above the Shannon oversampling rate of 2.5. 
In order to apply the Wiener filtering procedure, we need to 
compute the matrices $\expec{\vy\vy^t}$
and $\expec{\vx\vy^t}$. 
The former 
contains the correlation between the data points and themselves, 
and is given by
\beq{yytEq}
\expec{\y_i\y_j} = 
\expec{\n_i\n_j} + 
\int\int f_i(\rh)f_j(\rh')c(\rh\cdot\rh')d\Omega d\Omega',
\eeq
where the correlation function $c$ is given by the angular
power spectrum $C_\l$ through the familiar relation
\beq{CorrelationEq}
c(\cos\theta) = \sum_{\l=0}^{\infty} 
\left({2\l+1\over 4\pi}\right) P_\l(\cos\theta) C_\l,
\eeq
where $\P_\l$ are the Legendre polynomials. 
As described in N97, the noise covariance matrix 
$\expec{\vn\vn^t}$ is almost
diagonal for the Saskatoon experiment, 
but there are a small number of non-zero correlations 
(for example, there is a $2\%$ anticorrelation
between the Ka94 5pt East data and the simultaneously acquired Ka94 7pt East 
data due to mainly to atmospheric noise).

Likewise, the correlation between the data points and the pixels is given by
\beq{xytEq}
\expec{\x_i\y_j} = 
\int\int \pix(\rh_i\cdot\rh')f_j(\rh)c(\rh\cdot\rh')d\Omega d\Omega'.
\eeq

\subsection{Practical Issues}

To compute the covariance matrices
$\expec{\vx\vy^t}$ and $\expec{\vy\vy^t}$, we approximate 
the integrals in equations\eqnum{yytEq} and\eqnum{xytEq} by sums
over a $256\times 265$ grid of points.
Direct computation of $\expec{\vy\vy^t}$ with this procedure 
would take over a decade on a typical workstation, even if the 
code were optimized by omitting from the double sum all 
pixels where $\f_i$ or $f_j$ are zero.
Fortunately, the relevant angular separations are 
all much less the a radian, which means that
the effect of sky curvature is negligible. 
To a good approximation, 
we can thus integrate over a flat two-dimensional plane instead
and obtain
\beq{plane_yyEq}
\expec{\vy\vy^t} - \expec{\vn\vn^t} \approx
\int\int\f_i(\vr)\f_j(\vr') c(\vr-\vr')d^2r d^2r'
= \int\int\f_i(\vr)(\f_j\star c)(\vr)d^2r,
\eeq
where $\star$ denotes convolution.
Using fast Fourier transforms (FFTs) to compute 
the convolutions, the entire covariance matrix can now be computed
in merely a day. The computation of $\expec{\vx\vy^t}$ can be
accelerated in the same way.

Because the electronic offset is unknown, the means must be removed from 
the observations with each of the 64 synthesized beam patterns, 
which corresponds to multiplying the data vector $\vy$ by a 
certain projection matrix $\P$. 
We thus use the corrected covariance matrices
$\expec{\vx\vy^t}\P^t$ and $\P\expec{\vy\vy^t}\P^t$ 
in place of $\expec{\vx\vy^t}$ and $\expec{\vy\vy^t}$ 
in \eq{WienerEq}. 
We find that this correction makes a difference of only
a few percent.

\section{RESULTS AND CONCLUSIONS}

\begin{figure}[phbt]
\vskip-2cm
%\hglue2.5cm\hbox{\it Tegmark, de Oliveira-Costa, Devlin, Netterfield, Page \& Wollack 1996}\\
\centerline{\it Tegmark, de Oliveira-Costa, Devlin, Netterfield, Page \& Wollack 1996}
\centerline{\rotate[r]{\vbox{\epsfysize=16cm\epsfbox{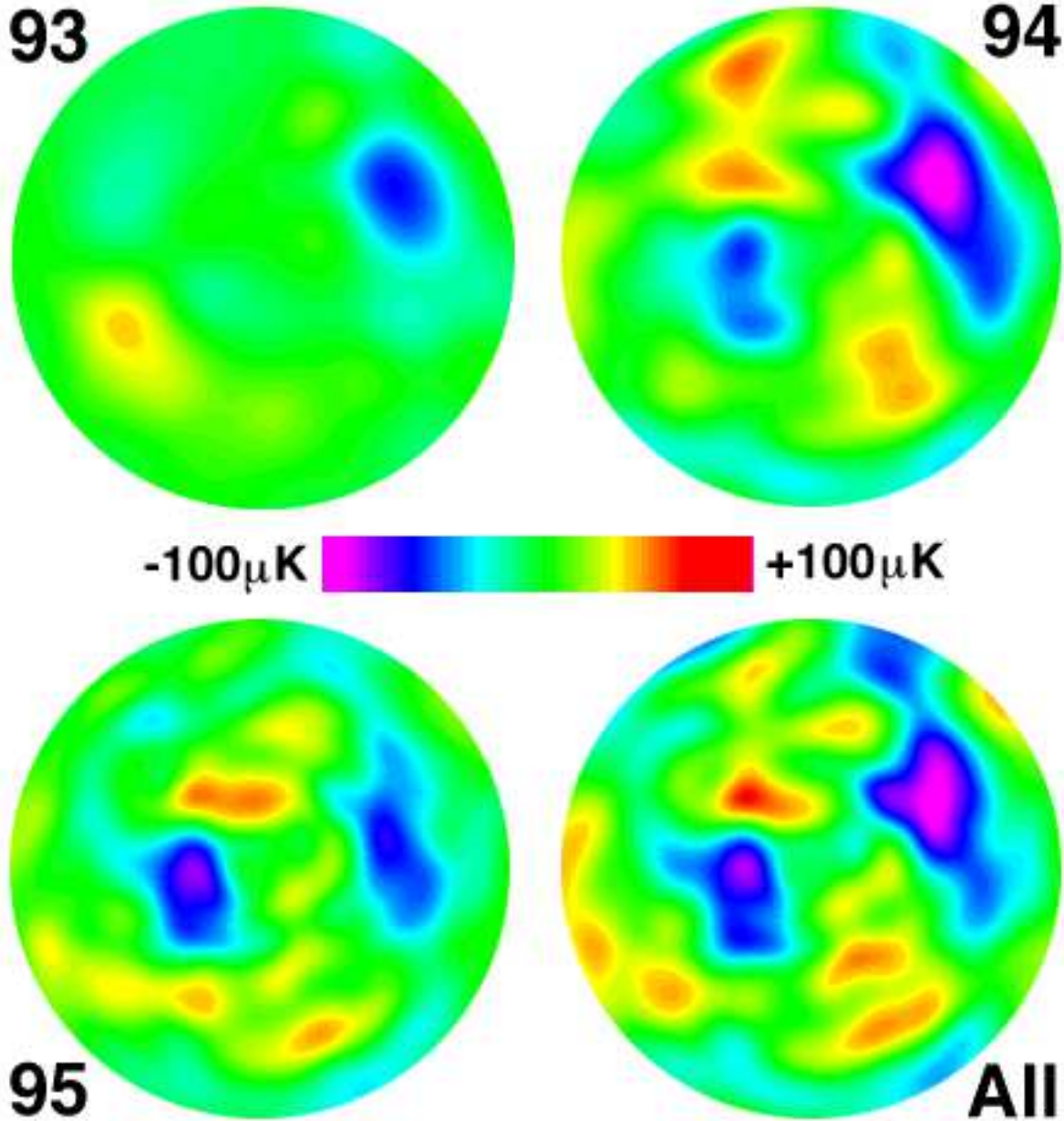}}}}

\caption{
Wiener-filtered maps.
}
The CMB temperature is shown in coordinates where the North Celestial Pole
is at the center of a circle of $16^\circ$ diameter, 
with $RA$ being zero at the top and increasing clockwise.
The first three panels show the maps 
using only the 1993, 1994 and 1995 data sets, respectively.
The last panel (bottom right) shows the map based on all
three years of data.
\label{MapFig}
\end{figure}

The resulting map is shown in \fig{MapFig} (bottom right)
and in \fig{3Dfig}. 
The fiducial power spectrum used is described in
section~\ref{DetailsSec} below. 
As expected, it contains virtually no features more than
$8^\circ$ from the center, 
reflecting the fact that the sky outside of this circle was not
probed by any of the beam functions. Computation of the
relevant covariance matrices shows that 
the signal-to-noise ratio within this disc is fairly constant and
of order two. In other words, the main features visible in this 
map are expected to be real rather than mere noise fluctuations.
We also generated a number of mock Saskatoon data sets, ran them 
through the inversion software and compared the reconstructions with the
original maps, which confirmed this conclusion.

\subsection{Comparison between Years}

The three first  panels in \fig{MapFig} show maps generated from
the subsets of the data that were taken 
in 1993, 1994 and 1995, respectively.
The 1993 data is seen to be rather featureless,
reflecting the fact that the 1993 data set 
(42 data points) contains considerably less information
than the other two years of data.
Similarly, the 1995 map is seen to contain more small-scale structure
than the 1994 map, reflecting the fact that the angular resolution
was approximately doubled in 1995.
Most potential sources of problems with the experiment
(underestimation of atmospheric contamination, sidelobe
pickup from celestial bodies, \etc) would be expected to vary on 
timescales much shorter than a year. In addition, 
the beam patterns were quite different in the three years
as described
in N97 -- for instance, the beam width was substantially reduced
in 1995 as mentioned above.
The visual similarity between these independent maps therefore 
provides reassuring evidence that the bulk of the signal being 
detected is in fact due to temperature-fluctuations on the sky rather 
than unknown systematic problems.

In addition to a qualitative visual inspection, the
1994 and 1995 maps
can be used to make more quantitative consistency checks. For
example, we subtracted one from the other and compared the resulting
noise levels with the theoretical expectations 
(as given by the diagonal of 
$W\expec{\vn\vn^t} W^t)$.
The levels are in good agreement, indicating
that there is no evidence for additional unmodeled/overlooked
sources of noise.

The perhaps most striking single feature in the maps, 
which stands out in all three years of data, 
is the large cold spot around ``two o'clock", about half
way from the center. It is interesting to note that the
existence of this cold spot can be
qualitatively inferred from the plots of the raw data 
in Wollack {\etal} (1993) and Netterfield {\etal} (1994), 
since the three-point beam (which has a positive lobe 
half-way from the center) was found to give large negative 
temperature between 4 and 5 hours in $RA$.

\subsection{Dependence on method details}
\label{DetailsSec}

To test if the reconstructed map is sensitive to the pixel size,
the analysis was repeated with $32\times 32$ pixels.
As expected, this produced virtually identical maps, since the
original $64\times 64$ pixel map was substantially oversampled.

The maps in \fig{MapFig} where generated using a 
a featureless (flat) fiducial power spectrum 
$C_\l=6Q^2/\l(\l+1)$ normalized to $Q=20\mK$.
To what extent do the maps depend on this choice?
As described below, the short answer to this question is
``almost not at all". 
As a test, we repeated the analysis for flat power spectra with
$Q=0$, $10\mK$, $47\mK$ (the best fit to the
Saskatoon power spectrum points of N97)
and $60\mK$, as well as for four different
normalizations of the standard CDM models 
(Sugiyama 1995). The spatial features remained
essentially unchanged, and the different normalizations 
simply caused different degrees of smoothing. 
The appearance of the map depended essentially only on 
one single property of the power spectrum: 
the broad-band power on the angular scales where the SK 
experiment is sensitive. 
This behavior is easy to understand
from \eq{WienerEq}.
Note that whereas $\expec{\vy\vy^t}$ is a sum of two contributions,
one from signal and one from noise, 
$\expec{\vx\vy^t}$ depends only on the signal. 
Roughly speaking, $\W$ is thus of the form signal/(signal+noise).
In the extreme case of no signal ($Q=0$), 
the Wiener-filtered map thus 
becomes identically zero, since $\expec{\vx\vy^t}=0$. 
If we increase the assumed signal-to-noise ratio, 
generic components of $\W$ increase in magnitude, 
and the Wiener filtering process will attempt to recover 
more details in the map. 
Since the noise loosely speaking enters on smaller scales 
than the signal, assuming a lower signal-to-noise ratio will 
basically cause 
the filtering to suppress high frequencies more than low frequencies,
\ie, smooth the map more. 
In summary, using a fiducial power spectrum with the wrong 
amount of power in the SK band will produce a map with 
the same spatial features in the same locations, but simply smoothed
either more or less than what is optimal.

Which is the best fiducial power level to use? 
The answer to this question depends on our desired 
signal-to-noise ratio $S/N$
(which we define as the ratio of the {\rms} signal and 
the {\rms} noise). The variance in a map pixel is given 
by the corresponding diagonal element of $W\expec{\vy\vy^t}W^t$, so 
we can separate the contributions
of signal and noise by splitting $\expec{\vy\vy^t}$ into a signal 
and a noise part and then compute $S/N$.  
Since the Wiener filtering
balances between smoothing too little (getting swamped by noise)
and smoothing too much (loosing unnecessarily much of the 
small-scale signal), it typically produces a map where these two 
problems are comparable in magnitude, {\ie}, where the noise 
is comparable to the {\it lost} part of the signal. 
Since $S/N$ compares the noise to the part of the signal which was
{\it not} lost, there is no a priori guarantee that the $S/N$ obtained 
will be satisfactory. It is thus common to adjust the fiducial 
power level to obtain a desired $S/N$. In our case, 
$S/N\approx 1.3$ for the combined map when the fiducial band power 
was $Q=47\mK$, so we chose $Q=20\mK$ to 
get a more smoothed and less noisy map, which has $S/N\approx 2.0$.

By dividing the map by the {\rms} noise $\sigma$, 
we can read off the significance
level of individual map features. 
\footnote{More generally, \eq{ErrCovEq} 
can be used to place error bars on linear combinations
of map pixels (such as correlations with external templates) 
and to make so-called constrained realizations 
(see {\eg} Zaroubi {\etal} 1995).
}
For instance, the 
cold spot around ``two o'clock" is $-7\sigma$, 
the one at ``eight o'clock" is $-5\sigma$ and
the hot spot at ``ten o'clock", near the center, 
is $+5\sigma$. 

\bigskip
In conclusion, we have presented the largest map to date of the 
CMB at degree scale angular resolution. The signal-to-noise ratio 
is of order two, and some individual hot and cold spots are significant
at the $5\sigma$ level. It is hoped that this map can be used to make
comparisons between experiments and with various foreground
templates, thereby improving our understanding of systematics and
foregrounds in preparation for the  
next generation of CMB missions.

\bigskip
The authors wish to thank David Wilkinson for helpful
comments on the manuscript.
Support for this work was provided by
NASA through a Hubble Fellowship,
HF-01084.01-96A, awarded by the Space Telescope Science
Institute, which is operated by AURA, Inc. under NASA
contract NAS5-26555, by European Union contract
CHRX-CT93-0120, by Deutsche Forschungsgemeinschaft grant
SFB-375, by NSF grant PH 89-21378, NASA grants NAGW-2801
and NAGW-1482, a Cottrell Scholar Award of Research Corp, a
David Lucile Packard Foundation Fellowship, and an NSF NII grant
to L. Page.

%%%%%%%%%%%%%%%%%%%%%% REFERENCES: %%%%%%%%%%%%%%%%%%%%%%%%%

%\clearpage

\section{REFERENCES}

\rf Bennett, C. L. 1996;ApJ;464;L1
% preprint astro-ph/9601067.
% Title: 4-Year COBE DMR Cosmic Microwave Background Observations: Maps and Basic Results 
% Author(s): C. L. Bennett , A. Banday , K. M. Gorski , G. Hinshaw , P. Jackson , P. Keegstra ,
% A. Kogut , G. F. Smoot , D. T. Wilkinson , E. L. Wright 
% ASTROPHYSICAL JOURNAL, 1996 JUN 10, V464 N1:L1+.

\rf Bond, J. R. 1995;Phys. Rev. Lett.;74;4369
% preprint astro-ph/9407044
% SIGNAL-TO-NOISE EIGENMODE ANALYSIS OF THE TWO-YEAR COBE MAPS.
% PHYSICAL REVIEW LETTERS, 1995 MAY 29, V74 N22:4369-4372.

\rf Bunn, E. F. {\etal} 1994;ApJ;432;L75
% Bunn, E.F.; Fisher, K.B.; Hoffman, Y.; Lahav, O.; and others.
% Wiener filtering of the COBE differential microwave radiometer data.
% Astrophysical Journal, Letters, 10 Sept. 1994, vol.432, (no.2, pt.2):L75-8.

\rf Bunn, E. F., Hoffmann, Y \& Silk, J 1996;ApJ;464;1
% preprint astro-ph/9509045.
% BUNN EF; HOFFMAN Y; SILK J.
% THE WIENER-FILTERED COBE DMR DATA AND PREDICTIONS FOR THE TENERIFE
% EXPERIMENT.
% ASTROPHYSICAL JOURNAL, 1996 JUN 10, V464 N1:1+.     
% Title: The Wiener-Filtered COBE DMR Data and Predictions for the Tenerife Experiment 
% Author(s): Emory F. Bunn , Yehuda Hoffman , Joseph Silk 

\rf Fisher, K. B. {\etal} 1995;MNRAS;272;885
% Fisher, K.B.; Lahav, O.; Hoffman, Y.; Lynden-Bell, D.; and others.
% Wiener reconstruction of density, velocity and potential fields from
% all-sky galaxy redshift surveys.
% Monthly Notices of the Royal Astronomical Society, 
% 15 Feb. 1995, vol.272, (no.4):885-908.

\rn Hu, W., Sugiyama, N. \& Silk, J. 1996, preprint astro-ph/9604166.

\rf Jungman, G.. Kamionkowski, M., Kosowsky, A \& 
Spergel, D. N. 1996;Phys. Rev. D;54;1332
% preprint astro-ph/9512139.
% JUNGMAN G; KAMIONKOWSKI M; KOSOWSKY A; SPERGEL DN.
% COSMOLOGICAL-PARAMETER DETERMINATION WITH MICROWAVE BACKGROUND MAPS.
% PHYSICAL REVIEW D, 1996 JUL 15, V54 N2:1332-1344.

\rf Lahav, O. {\etal} 1994;ApJ;423;L93
% Lahav, O.; Fisher, K.B.; Hoffman, Y.; Scharf, C.A.; and others.
% Wiener reconstruction of all-sky galaxy surveys in spherical harmonics.
% Astrophysical Journal, Letters, 10 March 1994, vol.423, (no.2, pt.2):L93-6.
    
\rf Netterfield {\etal} 1995;ApJ;445;L69
% NETTERFIELD CB; JAROSIK N; PAGE L; WILKINSON D; and others.
% THE ANISOTROPY IN THE COSMIC MICROWAVE BACKGROUND AT DEGREE ANGULAR
% SCALES. ASTROPHYSICAL JOURNAL, 1995 JUN 1, V445 N2:L69-L72.
     
\rf Netterfield, C. B. {\etal} 1997;ApJ;474;47
% preprint astro-ph/9601197.
% Title: A Measurement of the Angular Power Spectrum of the Anisotropy in the Cosmic
% Microwave Background 
% Author(s): C. B. Netterfield (Princeton) , M. J. Devlin (Princeton) , N. Jarosik (Princeton) , L.
% Page (Princeton) , E. J. Wollack (NRAO) 

\rf Rybicki, G. B. \& Press, W. H. 1992;ApJ;398;169
% Rybicki, G.B.; Press, W.H.
% Interpolation, realization, and reconstruction of noisy, irregularly
% sampled data.
% Astrophysical Journal, 10 Oct. 1992, vol.398, (no.1, pt.1):169-76.
% ONE-DIMENSIONAL CASE ONLY.
% They also have a 1995 PRL where they show that you can make stuff 
% tridiagonal if the 1D power spectrum is Lorentzian.

\rf Smoot, G. F. {\etal} 1992;ApJ;396;L1
% \rf Smoot, G.F.; Bennett, C.L.; Kogut, A.; Wright, E.L.; and others.
% Structure in the COBE Differential Microwave Radiometer first-year maps.

\rf Sugiyama, N. 1995;ApJS;100;281
% COSMIC BACKGROUND ANISOTROPIES IN COLD DARK MATTER COSMOLOGY.
% ASTROPHYSICAL JOURNAL SUPPLEMENT SERIES, 1995 OCT, V100 N2:281-305.
% The data paper.
     
\rf White, M. \& Bunn, E.F. 1995;ApJ;443;L53
% A first map of the cosmic microwave background at 0.5 degrees resolution.
% Astrophysical Journal, Letters, 20 April 1995, vol.443, (no.2, pt.2):L53-6.

\rn Wiener, N. 1949, {\it Extrapolation and Smoothing of
Stationary Time Series} (NY: Wiley). 

\rf Wollack, E. J. {\etal} 1993;ApJ;419;L49
% WOLLACK EJ; JAROSIK NC; NETTERFIELD CB; PAGE LA; and others.
% A MEASUREMENT OF THE ANISOTROPY IN THE COSMIC MICROWAVE BACKGROUND
% RADIATION AT DEGREE ANGULAR SCALES.
% ASTROPHYSICAL JOURNAL, 1993 DEC 20, V419 N2:L49-L52.
 
\rn Wollack, E. J. {\etal} 1996, preprint astro-ph/9601196.
% Title: An Instrument For Investigation of the Cosmic Microwave Background Radiation at
% Intermediate Angular Scales 
% Author(s): E. J. Wollack (NRAO) , M. J. Devlin (Princeton) , N. Jarosik (Princeton) , C. B.
% Netterfield (Princeton) , L. Page (Princeton) , D. Wilkinson (Princeton) 

\rf Zaroubi, S. {\etal} 1995;ApJ;449;446
% preprint astro/ph 9410080.
% Zaroubi, S.; Hoffman, Y.; Fisher, K.B.; Lahav, O.
% Wiener reconstruction of the large-scale structure.
% Astrophysical Journal, 20 Aug. 1995, vol.449, (no.2, pt.1):446-59.

%%%%%%%%%%%%%%%%%%%%%% FIGURES: %%%%%%%%%%%%%%%%%%%%%%%%%
 
\def\fheight{10.3cm} \def\fwidth{14.5cm}
 
\clearpage

\clearpage
\pagestyle{empty}

\clearpage
\pagestyle{plain}
\begin{figure}[phbt]
\rotate[l]{\vbox{\epsfysize=14.0cm\epsfbox{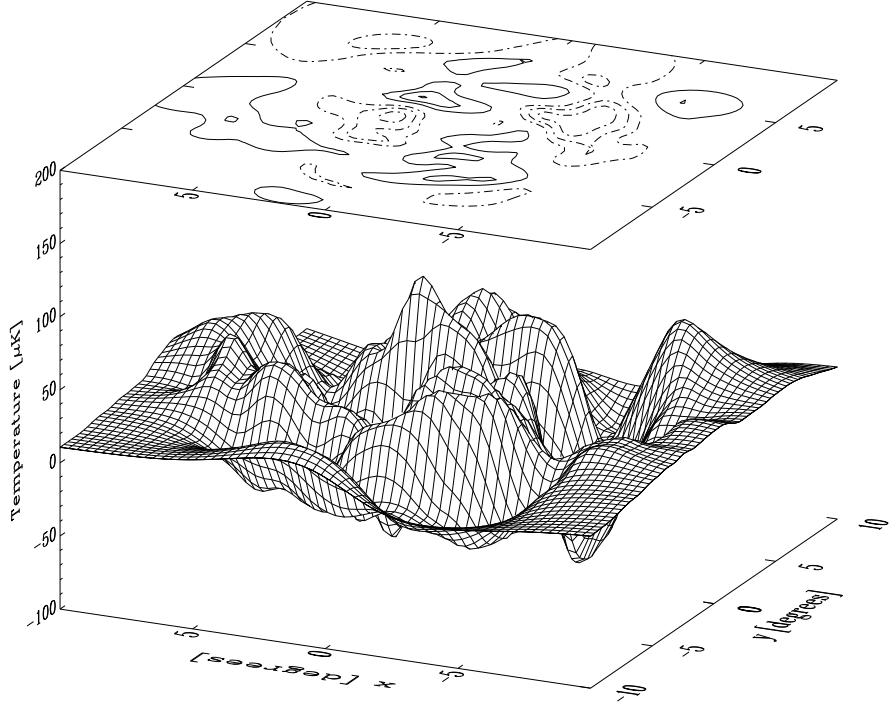}}}

\caption{
The Wiener-filtered map.
}
The map based on the entire data set (bottom right in 
\fig{MapFig}) is 
shown in coordinates where the North Celestial Pole
is at the center of a $20^\circ\times 20^\circ$ square  
with $RA$ being zero at the top and increasing clockwise.
$RA=0$ is at $(x,y) = (0,10^\circ)$.
The contour curves  correspond to 
$-75\mK$, $-50\mK$, $-25\mK$, $25\mK$, 
$50\mK$ and $75\mK$, respectively.
\label{3Dfig}
\end{figure}

\ed

\end{document}